\documentclass[letterpaper]{article}
\usepackage{aaaiB}
\usepackage{fixbib}
\usepackage{times}
\usepackage{helvet}
\usepackage{courier}
\usepackage{algorithm2e}

\usepackage{enumerate}
\usepackage{pdfsync}

\usepackage{url}

\usepackage{graphicx} 
\usepackage{textcomp} 
\usepackage{amsfonts} 
\usepackage{amsmath}
\usepackage{amssymb}
\usepackage{latexsym}
\usepackage{array}
\usepackage{float}
\usepackage{amsthm} 
\usepackage{calc}
\usepackage{perpage}
\usepackage{textcomp}
\usepackage{verbatim} 
\usepackage{marvosym}
\usepackage{hieroglf}
\usepackage{multirow} 
\usepackage{rotating} 
\usepackage{dsfont} 

\usepackage[english]{babel}

\usepackage{tikz}

\newtheorem{theorem}{Theorem}
\newtheorem{lemma}{Lemma}

\newcommand{\argmin}{\operatornamewithlimits{argmin}}

\def\R{\mathbb{R}}

\newcommand{\rp}{\mathbb{R}^p}
\newcommand{\rn}{\mathbb{R}^n}
\newcommand{\rnp}{\mathbb{R}^{n\times p}}

\newcommand{\otp}{\{1,\dots,p\}}

\newcommand{\be}{\widehat\beta}

\newcommand{\bt}{\beta^*}
\newcommand{\noise}{\epsilon}

\newcommand{\btrex}{\widehat{S}_{\text{B-TREX}}}

\graphicspath{ {./figures/} }

\usepackage[lofdepth,lotdepth, caption=false]{subfig}
\usepackage{dsfont}

\begin{document}
%
\title{Don't Fall for Tuning Parameters:\\Tuning-Free Variable Selection in High Dimensions With the TREX}


\author{
Johannes Lederer\\
Department of Statistical Science\\
Cornell University\\
Ithaca, NY 14853\\
\texttt{johanneslederer@cornell.edu}\\
\And
Christian L. M\"uller\\
NYU Courant Institute of Mathematical Sciences and\\
Center for Genomics and Systems Biology\\
New York, NY 10012\\
\texttt{cm192@nyu.edu}\\
}

\maketitle
\begin{abstract}
\begin{quote}
Lasso is a popular method for high-dimensional variable selection, but it hinges on a tuning parameter that is difficult to calibrate in practice.  In this study, we introduce TREX, an alternative to Lasso with an inherent calibration to all aspects of the model. This adaptation to the entire model renders TREX an estimator that does not require any calibration of tuning parameters. We show that TREX can outperform cross-validated Lasso in terms of variable selection and computational efficiency. We also introduce a bootstrapped version of TREX that can further improve variable selection. We illustrate the  promising performance of TREX both on synthetic data and on two biological data sets from the fields of genomics and proteomics.
\end{quote}
\end{abstract}

\section{Introduction}
In recent years, statistical tools that can deal with high-dimensional data and models have become pivotal in many areas of science and engineering. The advent of high-throughput technologies, for example, has transformed biology into a data-driven science that requires mathematical models with many variables. The need to analyze and reduce the complexity of these models has triggered an enormous interest in high-dimensional statistical methods that are able to separate relevant variables from irrelevant ones~\cite {BookAGS11,Buhlmann11,HastieTibshiraniFriedman2001}. Among the many existing methods, Lasso~\cite{Tibshirani:1996} and Square-Root Lasso (or Scaled Lasso)~\cite{Belloni11,Owen07,Stadler10,ScaledLasso11} have become very popular representatives.

In practice, however, high-dimensional variable selection turns out to be a difficult task. A major shortcoming of Lasso, in particular, is its need for a tuning parameter that is properly adjusted to all aspects of the model~\cite{YoyoMomo12} and therefore difficult to calibrate in practice. Using Cross-Validation to adjust the tuning parameter is not a satisfactory approach to this problem, because Cross-Validation is computationally inefficient and provides unsatisfactory variable selection performance. Replacing Lasso by Square-Root Lasso is also not a satisfactory approach, because Square-Root Lasso resolves only the adjustment of the tuning parameter to the variance of the noise but does not address the adjustment to the tail behavior of the noise and to the design. Similarly, more advanced Lasso-based procedures such as the Uncorrelated Lasso~\cite{Chen13} or the Trace Lasso~\cite{Grave11} also comprise tuning parameters that need proper calibration. In conclusion, none of the present approaches simultaneously provides parameter-free, accurate, and computationally attractive variable selection.

{\bf Our contribution:} In this study, we present a novel approach for high-dimensional variable selection. First, we reveal how a systematic development of the Square-Root Lasso approach leads to TREX, an estimator without any tuning parameter.  For optimal variable selection, we then combine TREX with a bootstrapping scheme. Next, we detail on implementations and demonstrate in a thorough numerical study that TREX is both accurate and computationally efficient. Finally, we discuss the findings and indicate directions for subsequent studies.
\section{Methodology}\label{theory}
 
\subsection{Framework for our study}

In this study, we aim at variable selection in linear regression. We therefore consider models of the form
\begin{equation}
  \label{d:model}
\tag{Model}
  Y= X\bt + \sigma\noise,
\end{equation}
where $Y\in\rn$ is a response vector, $X\in\rnp$ a design matrix, $\sigma>0$ a constant, and $\noise\in\rn$ a noise vector. We allow in particular for high-dimensional settings, where $p$ rivals or exceeds~$n$, and undisclosed distributions of the noise $\sigma\epsilon$. Statistical methods for models of the above form typically target $\bt$ (estimation), the support of $\bt$ (variable selection), $X\bt$ (prediction), or $\sigma^2$ (variance estimation). In this study, we focus on variable selection. 

To ease the exposition of the sequel, we append some conventions and notation: We allow for fixed and for random design matrices~$X$ but assume in either case the normalization $\left(X^\top X\right)_{jj}=n$  for all $j\in\otp$. Moreover, we assume that the distribution of the noise vector~$\epsilon$ has variance~$1$ so that $\sigma$ is the standard deviation of the entire noise $\sigma\epsilon$. Finally, we denote  the support (the index set of the non-zero entries) of a vector~$v$ by $\operatorname{support}(v)$ and the $\ell_q-$norm and the maximum norm of~$v$  by $\|v\|_q$ and $\|v\|_\infty$, respectively.

\subsection{TREX and B-TREX}

We now introduce two novel estimators for high-dimensional linear regression: TREX and B-TREX. To motivate these estimators, let us first detail on the calibration of Lasso. Recall that for a fixed tuning parameter $\lambda>0$, Lasso is a minimizer of a least-squares criterion with $\ell_1$-penalty:
\begin{equation}
\label{d:lasso}
\tag{Lasso}
  \be_{\text{Lasso}}(\lambda)\in\argmin_{\beta\in\rp}\left\{\frac{\|Y-X\beta\|_2^2}{n}+\lambda\|\beta\|_1\right\}.
\end{equation}
The tuning parameter $\lambda$ determines the intensity of the regularization and is therefore  highly influential, and it is well understood that a reasonable choice is of the order
\begin{equation*}
\lambda\sim \frac{\sigma\|X^\top \noise\|_\infty}{n}.  
\end{equation*}
For example, this becomes apparent when looking at the following prediction bound for Lasso (cf. \cite{Kolt10,RigTsy11}, see also~\cite{ArnakMoYo14} for an overview of Lasso prediction).
\begin{lemma}
  If $\lambda\geq 2{\sigma\|X^\top \noise\|_\infty}/n$, it holds
  \begin{equation*}
    \frac{\|X\be_{\text{\upshape Lasso}}(\lambda)-X\bt\|_2^2}{n}\leq 2\lambda\|\bt\|_1.
  \end{equation*}
\end{lemma}
This suggests a tuning parameter $\lambda$ that is small (since the bound is proportional to $\lambda$) but not too small (to satisfy the condition $\lambda\gtrsim {\sigma\|X^\top \noise\|_\infty}/n$). In practice, however, the corresponding calibration is very difficult, because it needs to incorporate several, often unknown, aspects of the model:\vspace{-2mm}

\begin{enumerate}[(a)]
\item the design matrix $X$;\vspace{-1mm}
 \item the standard deviation of the noise $\sigma$;\vspace{-1mm}
 \item the tail behavior of the noise vector $\noise$.
 \end{enumerate}

While one line of research approaches (a) and describes the calibration of Lasso to the design matrix~\cite{vdGeer11,YoyoMomo12,ArnakMoYo14}, Square-Root Lasso approaches (b) and eliminates the calibration to the standard deviation of the noise. To elucidate the latter approach, we first recall that for a fixed tuning parameter $\gamma>0$, Square-Root Lasso is defined similarly  as Lasso:
\begin{equation}\label{d:sqrtlasso}
\tag{Square-Root Lasso}
  \be_{\sqrt{\text{Lasso}}}(\gamma)\in\argmin_{\beta\in\rp}\left\{\frac{\|Y-X\beta\|_2}{\sqrt n}+\gamma\|\beta\|_1\right\}.
\end{equation}
Square-Root Lasso also requires a tuning parameter $\gamma$ to determine the intensity of the regularization. However, the tuning parameter should here be of the order (see, for example,~\cite{Belloni11})
\begin{equation*}
\gamma\sim \frac{\|X^\top \noise\|_\infty}{n},  
\end{equation*}
so that Square-Root Lasso does not require a calibration to the standard deviation of the noise. The origin of this feature can be readily located: Reformulating the definition of Square-Root Lasso as
\begin{equation*}
  \be_{\sqrt{\text{Lasso}}}(\gamma)\in\argmin_{\beta\in\rp}\left\{\frac{\frac{\|Y-X\beta\|_2^2}{n}}{\frac{\|Y-X\beta\|_2}{\sqrt n}}+\gamma\|\beta\|_1\right\}
\end{equation*}
identifies the factor ${\|Y-X\beta\|_2}/{\sqrt n}$ in the denominator of the first term as the distinction to Lasso. This additional factor acts as an inherent estimator of the standard deviation of the noise $\sigma$ and makes therefore the calibration to $\sigma$ obsolete. On the other hand, Square-Root Lasso still contains a tuning parameter that needs to be adjusted to (a) the design matrix and (c) the tail behavior of the noise vector.

We now develop the Square-Root Lasso approach further to address all aspects (a), (b), and (c). For this, we aim at incorporating an inherent estimation not of $\sigma$ but rather of the entire quantity of interest ${\sigma\|X^\top \noise\|_\infty}/{n}$. For this, note that if $\widehat \beta$ is a consistent estimator of $\bt$, then ${\sigma\|X^\top(Y-X\widehat\beta)\|_\infty}/{n}$ is a consistent estimator of ${\sigma\|X^\top\epsilon\|_\infty}/{n}$. In this spirit, we  define TREX\footnote{We call this new approach TREX to emphasize that it aims at {\it T}uning-free {\it R}egression that adapts to the {\it E}ntire noise $\sigma\epsilon$ and the design matrix $X$.} according to
\begin{align}
  \label{d:trex}\nonumber
  \be_{\text{TREX}}\in\argmin_{\beta\in\rp}\left\{\frac{\|Y-X\beta\|_2^2}{\frac 1 2\|X^\top (Y-X\beta)\|_\infty}+\|\beta\|_1\right\}\tag{TREX}.
\end{align}
Square-Root Lasso and Lasso are equivalent families of estimators (there is a one-to-one mapping between the tuning parameter paths of Square-Root Lasso and Lasso); in contrast, TREX is a {\it single, tuning-free} estimator, and its solution is in general not on the tuning parameter paths of Lasso and Square-Root Lasso. However, we can establish an interesting relationship between these paths and TREX (we omit all proofs for sake of brevity):
\begin{theorem}\label{thm}
It holds that
\begin{align*}
  &\min_{\beta\in\rp}\Bigg\{\frac{\|Y-X\beta\|_2^2}{\frac 1 2\|X^\top (Y-X\beta)\|_\infty}+\|\beta\|_1\\
&~~~~~~~~~~~~\text{~such that~}\|X^\top (Y-X\beta)\|_\infty\leq \|X^\top Y\|_\infty\Bigg\}\\
 =&\min_{0\leq u\leq 2\|X^\top Y\|_\infty/n}\Bigg\{\min_{\beta\in\rp}\Bigg\{\frac{\|Y-X\beta\|_2^2}{u}+\|\beta\|_1\\
&~~~~~~~~~~~~\text{~such that~}\frac 1 2\|X^\top (Y-X\beta)\|_\infty = u \Bigg\}\Bigg\}.
\end{align*}
\end{theorem}
\noindent In view of the Karush-Kuhn-Tucker conditions for Lasso, the latter formulation strongly resembles the Lasso path. This resemblance is no surprise: In fact, any consistent estimator $\widehat\beta$ of $\bt$ is related to a Lasso solution with an optimal (but in practice {\it unknown}) tuning parameter $\lambda\sim\sigma\|X^\top\epsilon\|_\infty/{n}$ via the formulation of TREX:
\begin{lemma}
  Assume that  $\widehat\beta$ a consistent estimator of $\bt$  and
\begin{equation*}
  \widetilde\beta\in\argmin_{\beta\in\rp}\left\{\frac{\|Y-X\beta\|_2^2}{\frac 1 2\|X^\top (Y-X\widehat\beta)\|_\infty}+\|\beta\|_1\right\}.
\end{equation*}
Then, $\widetilde\beta$ is close to a Lasso solution with tuning parameter $\lambda=\frac 1 2\|X^\top\epsilon\|_\infty/n$, that is,  
  \begin{equation*}
    \min_{\beta\in\Omega}\|\widetilde\beta-\beta\|_2=o(1)
  \end{equation*}
for $\Omega=\argmin_{\beta\in\rp}\left\{{\|Y-X\beta\|_2^2}+ \frac 1 2\|X^\top\epsilon\|_\infty\|\beta\|_1\right\}$.
\end{lemma}

Equipped with TREX to estimate the regression vector $\bt$, we can tackle a broad spectrum of tasks including estimation, prediction, and variance estimation. In this paper, however, we focus on variable selection. For this task,  we advocate an additional refinement based on sequential bootstrapping~\cite{Rao97}. More specifically, we advocate B-TREX for a fixed number of bootstraps $b\in\{1,2,\dots\}$:
\begin{algorithm}[h]
 \SetAlgoLined 
 \KwData{$(Y,X)$;}
 \KwResult{$\btrex\subset \otp$;}
\For{$i=1$ to $b$}{
Generate a sequential bootstrap sample $(\widetilde Y,\widetilde X)$;\\
Compute $\be_{\text{TREX}}$ on $(\widetilde Y,\widetilde X)$ according to~(TREX);\\
Set $\widehat S_i:=\operatorname{support}(\be_{\text{TREX}})$;
} 
Set $\btrex:=\{j:j\text{~is in more than half of the sets }\widehat S_1,\dots,\widehat S_b\}$;\\
 \caption{B-TREX with $b$ sequential bootstraps.}
\end{algorithm}\\
B-TREX is the majority vote over the TREX solutions for $b$ sequential bootstrap samples. Note that related bootstrapping schemes (based on traditional bootstrapping and different selection rules, however) have already been applied to Lasso~\cite{Bach08,FloriNeuro11}. In practice, it can also be illustrative to report the selection frequencies of each parameter over the bootstrap samples (cf. Figure~3). We finally note that B-TREX readily provides estimation and prediction if a least-squares refitting on the set $\btrex$ is performed. This refitting can improve the prediction and estimation accuracy if the set $\btrex$ is a good estimator of the true support of $\bt$~\cite{Belloni09,Lederer13}.

We point out that the norms $\|\cdot\|_\infty$ and $\|\cdot\|_1$ in the formulation of TREX are dual and that  extensions to other pairs of dual norms are straightforward.

A theoretical analysis of TREX is beyond the scope of this paper but is the subject of a forthcoming theory paper (with different authors). Note also that theoretical results for standard variable selection methods are incomplete: in particular, there are currently {\it no finite sample guarantees for approaches based on Lasso and Square-Root Lasso}: Finite sample bounds (``Oracle inequalities'')  for Lasso~\cite{Buhlmann11} and Square-Root Lasso~\cite{Yoyo13} require that the tuning parameters are properly calibrated to the model; yet, there are no guarantees that standard calibration schemes such as Cross-Validation or BIC-type criteria provide such tuning parameters.

\section{Implementation of TREX}\label{numerics}

To compute TREX, we consider the objective function 
\begin{equation*}
  \label{d:ftrex}\nonumber
  f_{\text{TREX}}:\beta \mapsto L(\beta)+\|\beta\|_1
\end{equation*}
that comprises the data-fitting term $L(\beta):=\frac{\|Y-X\beta\|_2^2}{\frac 1 2\|X^\top (Y-X\beta)\|_\infty}$ and the $\ell_1$-regularization term $\|\beta\|_1$. To make this objective function amenable to standard algorithms~\cite{Nesterov:2007,Schmidt:2010}, we invoke a smooth approximation of the data-fitting term. For this, we note that for all vectors $a\in\rp$ and positive integers $q\in\{1,2,\dots\}$, it holds that
\begin{equation*}
\|a\|_\infty\leq   \|a\|_q\leq p^{\frac 1 q}\|a\|_\infty,
\end{equation*}
and the data-fitting term $L(\beta)$ can therefore be approximated by the smooth data-fitting term
\begin{equation*}
\overline L(\beta) = \frac{\|Y-X\beta\|_2^2}{\frac 1 2\|X^\top (Y-X\beta)\|_q}.
\end{equation*}
We find that any $q \in [20, 100]$ works well in practice  (see supplementary material). We can calculate the gradient of the smooth approximation $\overline L(\beta)$ and obtain
\begin{align*}
&\nabla \overline L(\beta) =   \frac{2\|Y-X\beta\|_2^2 X^\top X (X^\top(Y-X \beta))^{q-1}}{\|X^\top (Y-X\beta)\|_q^{q+1}}\\
&~~~~~~~~~~~~~~~~~~~~~~~~~~~~~~~~~~~~~~~ - \frac{4X^\top(Y-X\beta)} {\|X^\top (Y-X\beta)\|_q}\, .
\end{align*}
The approximation $\overline L(\beta)+\|\beta\|_1$ of the criterion $f_{\text{TREX}}$ is now amenable to effective (local) optimization with projected scaled sub-gradient (PSS) algorithms~\cite{Schmidt:2010}. PSS schemes are specifically tailored to objective functions with smooth, possibly non-convex data-fitting terms and $\ell_1-$regularization terms. PSS algorithms only require zeroth- and first-order information about the objective function, have a linear time and space complexity per iteration, and are especially effective for problems with sparse solutions. Several PSS algorithms that fit our framework are described in~\cite[Chapter~2.3.1]{Schmidt:2010}\footnote{{\small http://www.di.ens.fr/\%7Emschmidt/Software/L1General.html} provides the implementations}. Among these algorithms, the Gafni-Bertsekas variant was particularly effective for our purposes.

The smooth formulation of the TREX criterion remains non-convex; therefore, convergence to the global minimum cannot be guaranteed. Neverthless, we show that the above implementation is fast, scalable, and provides estimators with excellent statistical performance.

Note also that the advent of novel optimization procedures~\cite{Breheny2011,Mazumder:2011} lead to an increasing popularity of non-convex regularization terms such as the Smoothly Clipped After Deviation (SCAD)~\cite{Fan:2001} and Minimax Concave Penality (MCP)~\cite{Zhang10}. More recently, also objective functions with non-convex data-fitting terms have been proved both statistically valuable and efficiently computable~\cite{Loh:2013,Nesterov:2007,Wang:2013}.

\section{Numerical Examples}\label{testcases}
We demonstrate the performance of TREX and B-TREX on three numerical examples. We first consider a synthetic example inspired by~\cite{Belloni11}. We then consider two high-dimensional biological data sets that involve riboflavin production in B.~subtilis~\cite{Buhlmann:2014} and mass spectrometry data from melanoma patients~\cite{mian05}. 

We perform the numerical computations in MATLAB 2012b on a standard MacBook Pro with dual 2GHz Intel Core i7 and 4GB 1333MHz DDR3 memory. To compute Lasso and its cross-validated version, we use the MATLAB-internal procedure \texttt{lasso.m} (with standard values), which follows the popular glmnet R code. To compute TREX, we use Schmidt's PSS algorithm implemented in \texttt{L1General2\_PSSgb.m} to optimize the approximate TREX objective function with $q=40$. We use the PSS algorithm with standard parameter settings and set the initial solution to the parsimonious all-zeros vector $\beta_\text{init} = (0,\dots,0)^\top\in\rp$. We use the following PSS stopping criteria: minimum relative progress tolerance $\text{optTol=1e-7}$, minimum gradient tolerance $\text{progTol=1e-9}$, and maximum number of iterations $\text{maxIter} = \max(0.2p,200)$. As standard for the number of bootstrap samples in B-TREX we set $b=31$. 

\subsection{Synthetic Example}\label{synthetic}
We first evaluate the scalability and the variable selection performance of TREX and B-TREX on synthetic data. The method of comparison is Lasso with the tuning parameter that leads to minimal $10-$fold cross-validated mean squared error (Lasso-CV). We generate data according to the linear regression model (Model) with parameters inspired by the Monte Carlo simulations in~\cite{Belloni11}: We set the sample size to $n = 100$, the number of variables to $p = 500$ (or vary over $p$), and the true regression vector to $\beta^{*} = (1,1,1,1,1,0,\ldots,0)^\top $; we sample standard normal errors $\epsilon \sim \mathcal{N}(0,\operatorname{I}_n)$ and multiply them by a fixed standard deviation $\sigma \in \{0.1,0.5,1,3\}$; and we sample the rows of $X$ from the $p-$dimensional normal distribution $\mathcal{N}(0,\Sigma)$, where $\Sigma$ is the covariance matrix with diagonal entries $\Sigma_{ii} =1$ and off-diagonal entries $\Sigma_{ij} = \kappa$ for $i,j\in\{1,\dots,p\}$ and a fixed correlation $\kappa \in \{0,0.5,0.9\}$, and then normalized them to Euclidean norm $\sqrt n$. We report scalability and variable selection results averaged over $51$ repetitions (thick, colored bars) and the corresponding standard deviations (thin, black bars). More precisely, we report the runtime of plain Lasso and of TREX as a function of $p$ (for $n=100$, $\sigma=0.5$, $\kappa=0$) in Figure~1, and we report the runtime and the variable selection performance of Lasso-CV, TREX, and B-TREX in Hamming distance for fixed $p=500$ in Figure~2.  

\begin{figure}[t]
\centering
\includegraphics[width=0.37\textwidth]{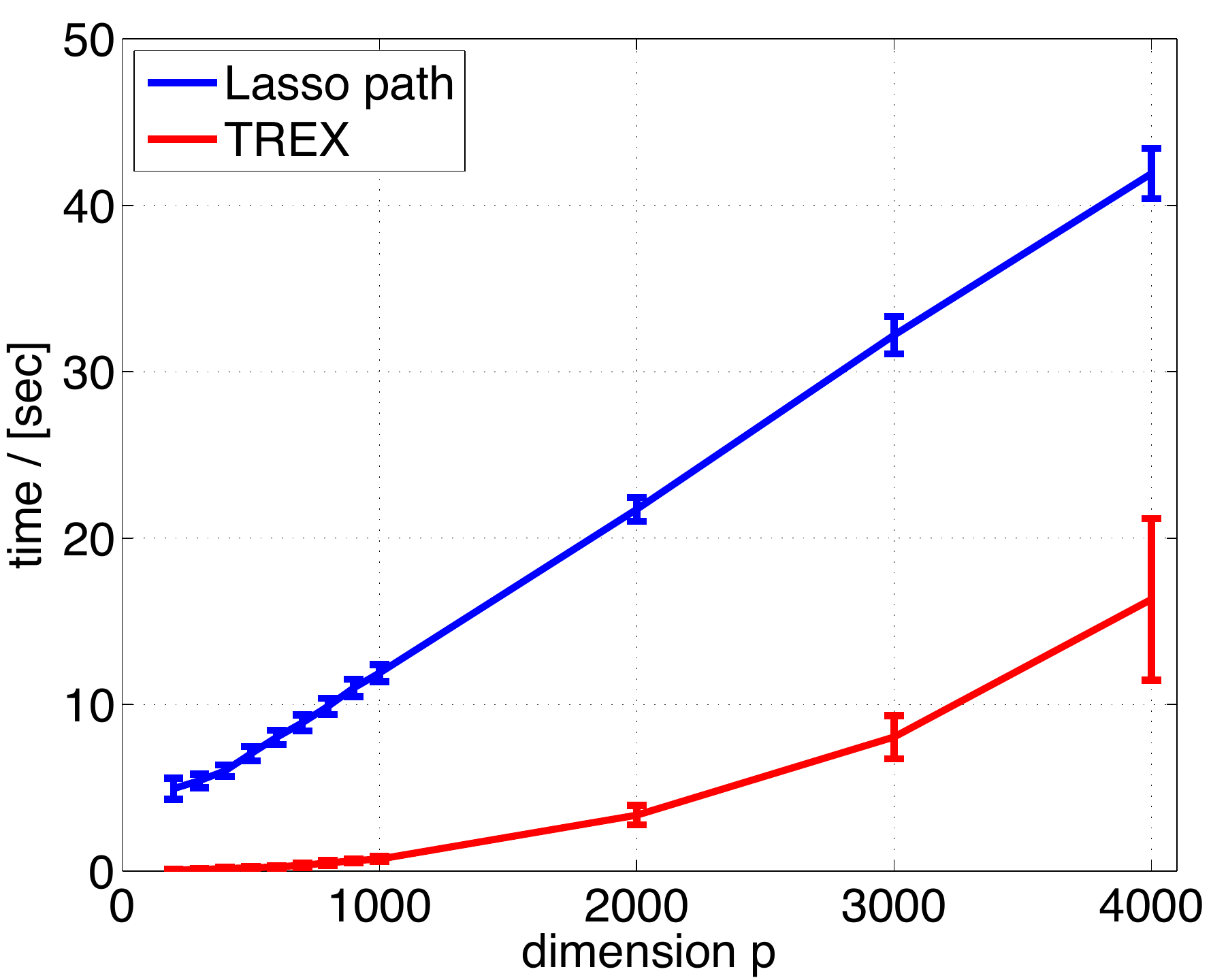}
\vspace{-0.9mm}
\caption{Runtime (in seconds) of TREX and {\it plain} Lasso as a function of $p$.}
\label{fig:synDatascale}
\end{figure}

The data shown in Figure~1 suggest that the runtime for TREX is between quadratic and cubic in $p$ (a least-squares fit results in $\mathcal O(p^{2.4})$) and, thus, illustrates the scalability of TREX at least up to $p=4000$. In comparison, the runtime for a single Lasso path (\textit{without} Cross-Validation or any other calibration scheme), shown in Figure~1, reveals a near-linear dependence of $p$ (a least-squares fit results in $\mathcal O(p^{1.1})$), though with a higher offset and slope.

Figure~2 summarizes the numerical results for the settings with $\kappa=0$. The runtimes disclosed in Figure~2 indicate that both TREX and B-TREX can rival Lasso-CV in terms of speed. The variable selection results show that TREX provides near-perfect variable selection for $\sigma\in\{0.1,0.5\}$ and B-TREX for $\sigma\in\{0.1,0.5,1\}$; for stronger noise, the Hamming distance of these two estimators to $\bt$ increases. Lasso-CV, on the other hand, consistently selects too many variables. For~$\kappa\in\{0.5, 0.9\}$ (see supplementary material), the performance of TREX deteriorates as compared to Lasso-CV. B-TREX, on the other hand, provides excellent variable selection for all considered parameter settings.  In summary, the numerical results for the standard synthetic example considered here provide first evidence that TREX and B-TREX can  outmatch Lasso-CV in terms of variable selection.

\begin{figure}[t]
\centering
\includegraphics[width=0.3\textwidth]{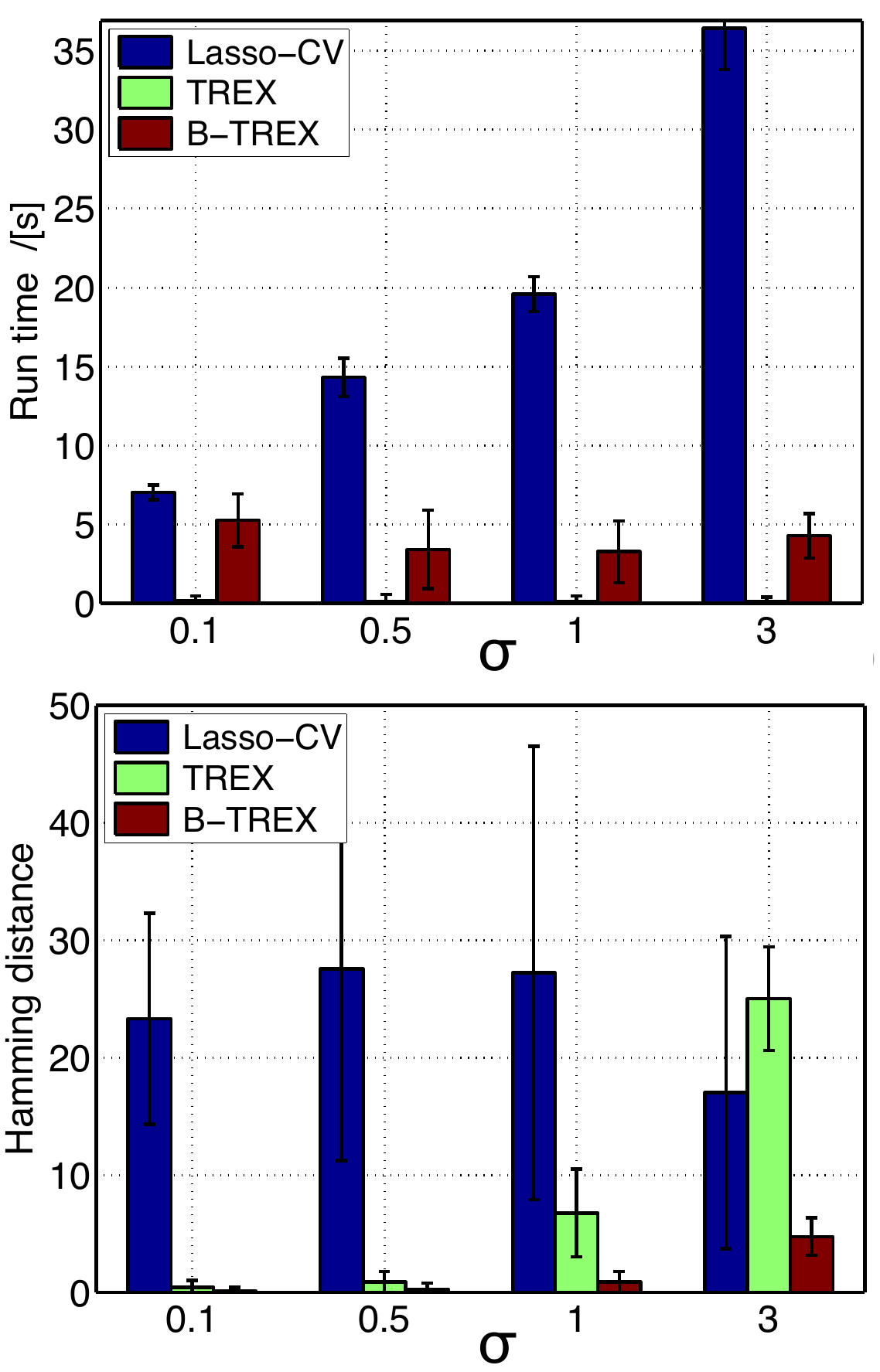}
\vspace{-0.9mm}
\caption{Runtimes (in seconds) and variable selection errors in Hamming distance on the synthetic example with $\kappa=0$ and $p=500$.}
\label{fig:synData}
\end{figure} 

\subsection{Riboflavin Production in B. Subtilis}\label{subtilis}
We next consider a recently published high-dimensional biological data set for the production of riboflavin (vitamin B${}_2$) in B. subtilis (Bacillus subtilis)~\cite{Buhlmann:2014}. The data set comprises expression profiles of $p=4088$ genes of different B. subtilis strains for a total of $n=71$ experiments with varying settings. The corresponding expression profiles are stored in the matrix $X\in\R^{71\times 4088}$. Along with these expression profiles, the associated standardized riboflavin log-production rates $Y\in\R^{71}$ have been measured. The main objective is now to identify a small set of genes that is highly predictive for the riboflavin production rate. 

We first report the outcomes of standard Lasso-based approaches, which can be obtained along the lines of~\cite{Buhlmann:2014}. The runtime for the computation of a single Lasso path with the MATLAB routine is approximately 58 seconds. Lasso-CV selects 38 genes, that is, its solution has 38 non-zero coefficients; the 20 genes with largest coefficients and the associated coefficient values are listed in Table~1. For variable selection, B\"uhlmann {\it et al.} specifically propose stability selection~\cite{Meinshausen:2010}. The standard stability selection approach is based on $500$ Lasso computations on subsamples of size $\lfloor{\frac{n}{2}}\rfloor$ and the $20$ coefficients that enter the corresponding Lasso paths first. This approach yields three genes: $\text{LYSC\_at}$, $\text{YOAB\_at}$, and $\text{YXLD\_at}$~\cite{Buhlmann:2014}. 

We next apply TREX and B-TREX. The runtime for a single TREX computation is approximately 30 seconds. TREX selects 20 genes and therefore provides a considerably sparser solution than Lasso-CV; the corresponding genes and the associated coefficients are listed in Table~1. B-TREX with the standard majority vote selects three genes: $\text{YXLE\_at}$, $\text{YOAB\_at}$, and $\text{YXLD\_at}$. The outcomes of B-TREX with  selection rules different from majority vote can be deduced from Table~1, where we list the selection frequencies of the 20 genes that are selected most frequently across the bootstraps.

The numerical results reveal three key insights: First, the set of genes selected by TREX and the set of the 20 genes corresponding to the highest coefficients in the Lasso-CV solution are distinct but share a common subset of $12$ genes. Second, the sets of genes selected by B-TREX and Lasso-CV stability selection have the two top-ranked Lasso-CV and TREX genes in common. On the other hand, the gene associated with the highest frequency in the B-TREX solution is not selected by stability selection. The B-TREX solution is biologically plausible: Since the genes  $\text{YXLD\_at}$ and $\text{YXLE\_at}$ are located in the same operon, both genes are likely to be co-expressed and involved in similar cellular functions. Third, the runtime for a single Lasso path is about two times larger than for a single TREX solution.

\begin{table}[t]
  \centering
  \includegraphics[width=0.45\textwidth]{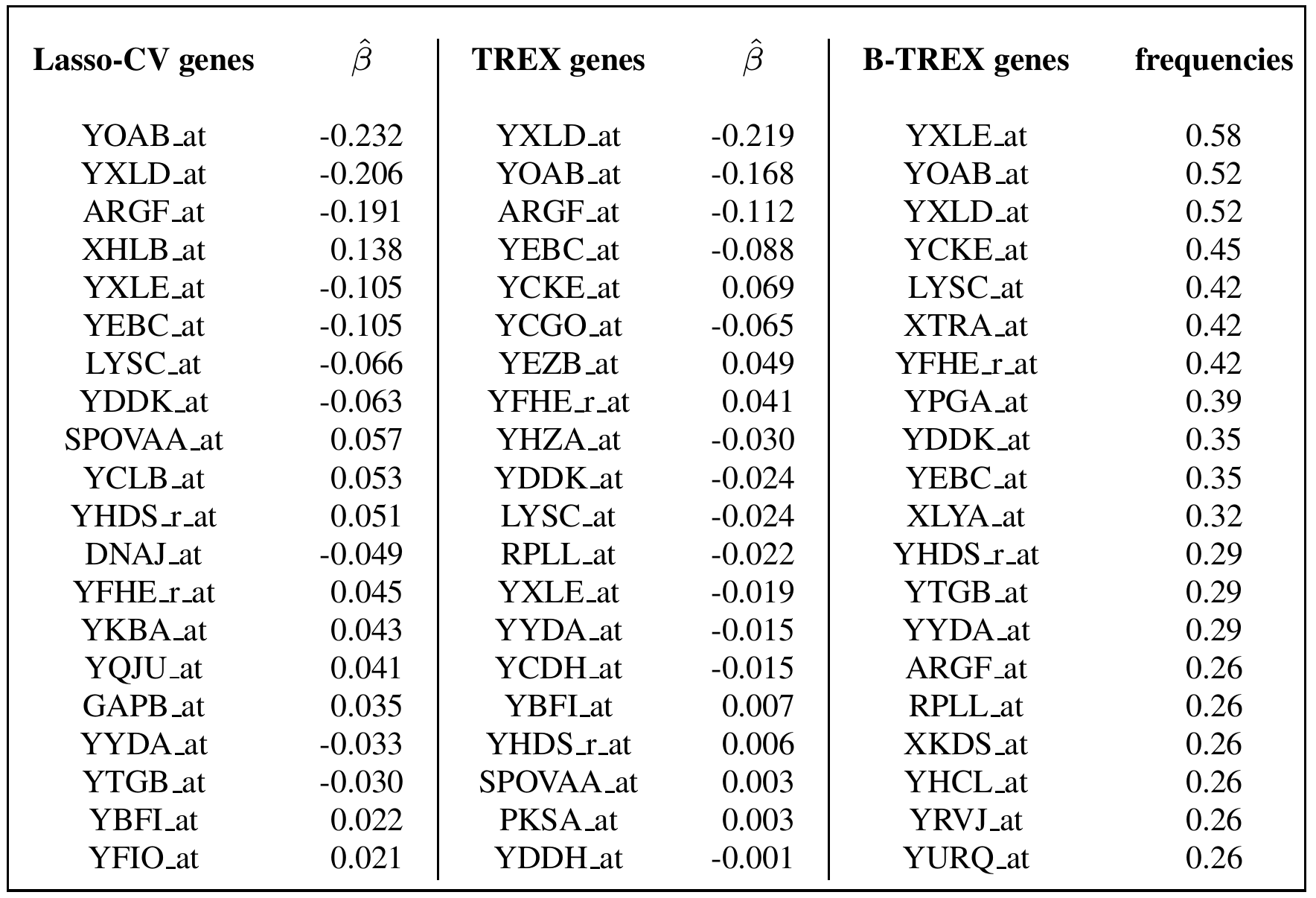}
\vspace{-0mm}
  \caption{Gene rankings for riboflavin production in B. subtilis. The left panel contains the 20~genes with largest coefficients in Lasso-CV (out of 38 genes with non-zero coefficients) and the associated coefficients. The center panel contains the 20 genes with non-zero coefficients in TREX and the associated coefficients. The right panel contains the 20 genes with largest frequencies in B-TREX and the associated frequencies.}
  \label{fitgenes}
\end{table}

The model complexities differ considerably, ranging from $3$ parameters for B-TREX to $38$ for Lasso-CV, and in applications, simple models are often preferred. We evaluate, therefore, the Leave-One-Out Cross-Validation errors (LOOCV-errors) of the methods under consideration for fixed numbers of parameters. As a reference, we report the LOOCV-errors of Lasso-CV (with the cross-validations performed on the training sets of size $n-1$) in the first row of Table~2. In the three subsequent rows, we then show the LOOCV-errors of TREX, of TREX with least-squares refitting (TREX-LS), and of Lasso with tuning parameter such that the number of non-zero entries equals the number of non-zero entries of TREX (Lasso-T). Finally, we give the LOOCV-errors of B-TREX and of Lasso with tuning parameter such that the number of non-zero entries equals the number of non-zero entries of B-TREX (Lasso-BT). The computations for Stability Selection are very intensive and therefore omitted. We observe that for fixed model complexity, the solutions of TREX (with least-squares refitting) and B-TREX have lower LOOCV-error than their Lasso-based counterparts. 

We conclude that the genes selected by B-TREX are commensurate with biological knowledge and that B-TREX can provide small models with good predictive performance.

\subsection{Classification of Melanoma Patients}
We also demonstrate the usefulness of the ranked B-TREX list for a proteomics data set from a study on melanoma patients~\cite{mian05}. The data\footnote{see \textcolor{black}{\small http://www.maths.nottingham.ac.uk/\%7Eild/mass-spec}} consist of $n=205$ mass spectrometry scans of serum samples from $101$~patients with Stage I melanoma (moderately severe) and $104$ patients with Stage IV melanoma (very severe). Each scan measures the intensities for $18\ 856$ mass over charge (m/Z) values. The objective is to find  m/Z values that are indicators for the stage of the disease, eventually leading to proteins that can serve as discriminative biomarkers~\cite{mian05}. 

We want to compare outcomes of our estimators with results described in~\cite{Vasilu:2014}. For this, we use the same linear regression framework (even though one could also argue in favor of a logistic regression framework) and the same data pre-processing: We apply an initial peak filtering step that yields the $p=500$ most relevant m/Z values. The resulting data are then normalized and stored in the matrix $X\in\R^{205\times 500}$. Next, the class labels in $Y\in\R^{205}$ are set to $Y_i = -1$ for $i=1,\ldots,101$ (Stage I patients) and to $Y_i = 1$ for $i=102,\ldots,205$  (Stage IV patients).
 
We now demonstrate the usefulness of the ranked list of predictors provided by B-TREX. For this, we first report in Figure~4 the parameter values of the least-squares refitted versions of the three estimators $10$-fold cross-validated Lasso (Lasso-CV), TREX, and B-TREX.  Lasso-CV selects $43$~predictors,  TREX selects $8$~predictors, and B-TREX selects $2$~predictors. We now use the signs of the (least-squares refitted) responses to estimate the class labels, cf.~\cite{Vasilu:2014}. We depict in Figure~4 averaged $10$-fold classification errors of Sure-Independence Screening (SIS), Iterative SIS (ISIS), Elastic Net, and Penalized Euclidean Distance (PED) (all taken from~\cite{Vasilu:2014}) and of TREX, B-TREX, and Lasso-CV. TREX shows almost identical classification error/model complexity as SIS and ISIS and outperforms Elastic net in terms of model complexity. PED and Lasso-CV have lower classification error but higher model complexity. B-TREX with standard majority vote results in a very sparse model with moderate error. More importantly, classification based on the top predictors from B-TREX is insensitive with respect to the threshold: For {\it any} number of predictors from $6$ up to $23$, B-TREX  outperforms {\it all} other estimators. We conclude that the ranked list of B-TREX predictors can lead to very robust and accurate model selection and,  in particular, can outperform on this data set all other standard estimators.

\begin{table}[t]
\centering
\begin{tabular}{l c c}
&~LOOCV-error~&\# of coefficients\\
\hline\hline Lasso-CV&0.42&39\\
\hline\hline
{TREX}&{0.51}&{21}\\
{TREX-LS}&{0.45}&{21}\\
Lasso-T&0.47&21\\
\hline\hline
{B-TREX}&{0.50}&{~~4}\\
Lasso-BT&0.62&~~4\\
\end{tabular}
\caption{Means of the Leave-One-Out Cross-Validation errors and median of the corresponding numbers of non-zero coefficients on the riboflavin dataset.\vspace{-6mm}}\label{tab:ribo}
\end{table}

\begin{figure}[t]
  \centering
  \includegraphics[width=0.35\textwidth]{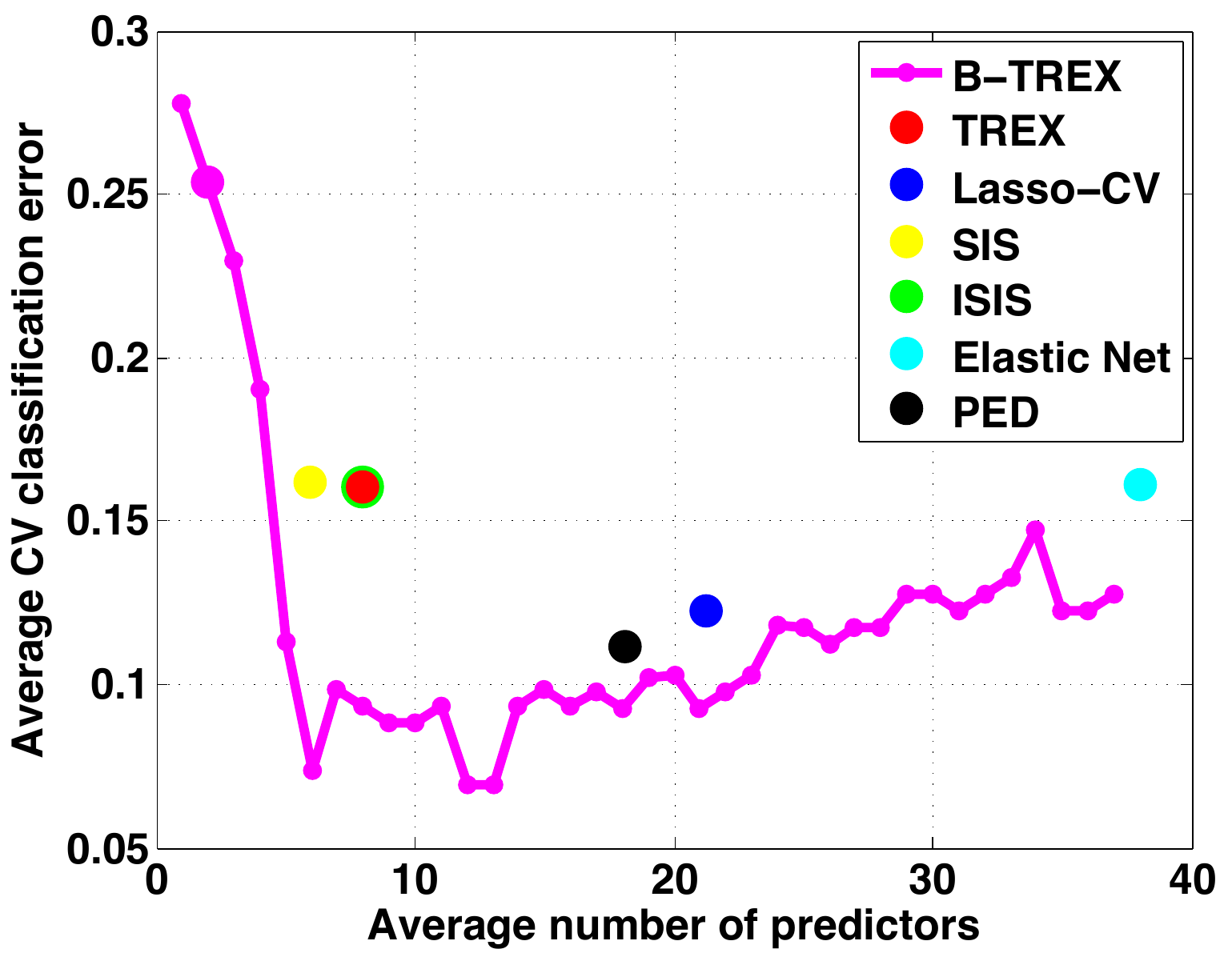}
\vspace{-0mm}
\caption{Mean 10-fold CV classification errors vs. average number of predictors.}
  \label{fig:fitmz}
\end{figure}

\section{Conclusions}\label{conclusions}
We have introduced TREX, a simple, fast, and accurate method for high-dimensional variable selection. We have shown that TREX avoids tuning parameters and, therefore, challenging calibrations. Moreover, we have shown  that TREX can outmatch a cross-validated Lasso in terms of speed and accuracy.

To further improve variable selection, we proposed B-TREX, a combination of TREX with a bootstrapping scheme. This proposition is supported by the numerical results and in line with earlier claims that bootstrapping can improve variable selection~\cite{Bach08,FloriNeuro11}. Moreover, we argue that the solution of B-TREX on the recent riboflavin data set in~\cite{Buhlmann:2014} is supported by biological insights. Finally, the results on the melanoma data show that TREX can yield robust classification.

Our contribution therefore suggests that TREX and B-TREX can challenge standard methods such as cross-validated Lasso and can be valuable in a wide range of applications.  We will provide further theoretical guarantees, optimized implementations, and tests for prediction and estimation performance in a forthcoming paper. A TREX MATLAB-toolbox as well as all presented numerical data will be made publicly available at the authors' websites.

\section*{Acknowledgments}
We sincerely thank the reviewers for their insightful comments and Jacob Bien, Richard Bonneau, and Irina Gaynanova for the valuable discussions.

\bibliographystyle{named}
\bibliography{Literature}

\end{document}